# The whole prefrontal cortex is premotor cortex

**Justin M. Fine and Benjamin Y. Hayden**

Department of Neuroscience,
Center for Magnetic Resonance Research,
and Department of Biomedical Engineering
University of Minnesota,
Minneapolis, MN 55455

**Correspondence**:
Benjamin Y. Hayden
CMRR Building,
University of Minnesota,
Minneapolis, MN 55455
Telephone: 425-749-2341
Email: benhayden@gmail.com

**Funding statement**
This research was supported by a National Institute on Drug Abuse Grant R01 DA038615 and MH124687 (to BYH).

**Competing interests**
The authors have no competing interests to declare.

**Acknowledgements**
We thank Sarah Heilbronner, Becket Ebitz, Maya Wang, and Michael Yoo for helpful conversations.

## ABSTRACT

We propose that the entirety of the prefrontal cortex can be seen as fundamentally premotor in nature. By this, we mean that the prefrontal cortex consists of an action abstraction hierarchy whose core function is the potentiation and depotentiation of possible action plans at different levels of granularity. We argue that the hierarchy's apex should revolve around the process of goal-selection, which we posit is inherently a form of abstract action optimization. Anatomical and functional evidence supports the idea that this hierarchy originates on the orbital surface of the brain and extends dorsally to motor cortex. Our view, therefore, positions the orbitofrontal cortex as the central site for the optimization of goal selection policies, and suggests that other proposed roles are aspects of this more general function. We conclude by proposing that the dynamical systems approach, which works well in motor systems, can be extended to the rest of prefrontal cortex. Our proposed perspective will reframe outstanding questions, open up new areas of inquiry, and will align theories of prefrontal function with evolutionary principles.

# MAIN TEXT

## Introduction

As we move around the world, our bodies engage in small movements that are often unrelated to the task at hand. An important recent study shows that, in mice, these small movements account for a large amount of the explainable variance in firing rate of neurons (Musall et al., 2019; see also related findings in Stringer et al., 2018 and Steinmetz et al., 2019). These effects were found not just in motor cortex, but, surprisingly, across the entire brain. These findings came about as a result of the careful measure and registration of the full suite of animal behavior; previous studies that did not measure these movements would have treated them as a source of noise to be ignored. Overall, these results highlight the importance of motor control for the brain as a whole.

It's fascinating how much of the neural response is determined by seemingly unimportant motor activity. Likewise, it is surprising the extent to which the cognitive variables that are central to so many models of cognition wind up being relatively small factors in determining the firing rates of neurons (Musall et al., 2019). Indeed, it's somewhat humbling to see that these cognitive variables - the focus of so much scholarship - are such a minor influence on firing rates. Despite decades of debate about how these regions differ functionally, when we consider factors that drive firing rates the most, these regions turn out to largely have the same function. This is not to say that these results support mass action theories. However, they invite us to ask whether studies that focus on differences in brain areas are ignoring the much larger common factors that drive all the regions.

Indeed, from another perspective, these results should not be *too* surprising. After all, the brain exists - first and foremost - to control behavior (Krakauer et al., 2017; Merel et al., 2019; Botvinick, 2008). From that perspective, the brain's other functions, including the ones that correspond to the chapters of any cognitive neuroscience textbook (attention, reward, memory, executive function, etc), are there to influence action. If they don't influence action, they are otiose. And if they do influence action, they are, essentially, modulatory factors for the expression of action. Still, what is still surprising is how large the motor effects are even in regions with no obvious motor role. That suggests that, when factoring in full motor behavior, so much of the brain has a motor or premotor role. In other words, these results advocate for the primacy of motor expression for understanding neural activity, not just in motor regions, but brainwide. They suggest, at least to us, that motor activity not only accounts for much of the variance in neural firing, but serves as the organizing structure for the rest of our mental activity. They also cohere with an evolution-centric view of functional neuroanatomy - anything that does not advance the cause of driving behavior should not last long as a major part of the brain's repertoire (Barrett, 2011).

**The whole prefrontal cortex is a premotor structure**

We believe that taking seriously the primacy of motor expression in driving brain activity has important implications for systems neuroscience. In particular, we think that this view can help organize understanding of the ever-mysterious prefrontal cortex (Badre and Nee, 2018; Christoff et al., 2009; Miller and Cohen, 2001; O'Reilly and Frank, 2006). This large portion of the brain is typically associated with non-motor cognitive processes, such as executive function and control, as well as working memory, inhibition, learning, and maintaining and switching task set (Badre and Nee, 2018; Mansouri et al., 2017; Fuster, 2015; Stuss and Knight, 2013; Miller and Cohen, 2000). Note that we are not arguing that other theories of PFC are incorrect. Our ideas, outlined below, will be speculative and are in need of data to support them. Moreover, we believe that full understanding of a structure as complex as PFC benefits from multiple perspectives - all of which, including our own, have limitations. However, we believe that these views *can* be felicitously augmented by considering the fundamental role of the PFC in driving or setting the stage for action, and seeing the proposed roles of its constituent regions through that lens.

This viewpoint is part of a larger view that advocates for thinking of cognition as an extension of action selection, not as something wholly separate from it (Clark, 1997; Klatzky et al., 2008; Pezzulo and Cisek, 2016; Cisek and Kalaska, 2010; Cisek, 2006; Cisek, 2012; Thelen et al., 2001). We humans (or other animals) move through the world and happen to come upon things that interest us. Those things that we encounter are associated with specific actions. From an economic perspective, the relevant action would be *selection*; in foraging theory, it would be *pursuit* or *handling* (Stephens and Krebs, 1986). In Gibsonian psychology, the roughly analogous concept is that we encounter options that activate an *affordance* associated with selection (Cisek, 2007; Cisek and Kalaska, 2010; Gibson, 1979; Turvey, 1992). Each potential action can either be performed or not performed. If a potential action rises to the level of consideration, the brain gathers available evidence, filters it for relevance, and uses that to militate for or against performing the action. The brain uses the same accept-reject principles for both trivial decisions and serious ones like choosing whether to buy a house or marry a partner (Kacelnik et al., 2011; Hayden, 2018; Hayden and Moreno-Bote, 2018; Kolling et al., 2016). ***The processes that increase or decrease the likelihood of performing an action wind up guiding the selection of actions, and are therefore – in a non-trivial sense - premotor.*** Because guiding these decisions is the chief function of the prefrontal cortex, the whole prefrontal cortex can validly be called premotor cortex.

**Wait, what is premotor cortex again?**

Premotor cortex, as the term is traditionally used, is defined by its relationship with motor cortex (Wise, 1985; Wise et al., 1997). Motor cortex is, of course, cortex whose chief function is the regulation either by planning, modifying, or executing movements, or some combination of those (Georgopoulos et al., 1986; Grazziano et al., 2002;  Omrani et al., 2016). Premotor cortex is thought to have a more abstract and high-level motor function, serving a more regulatory or supervisory role (Coallier et al., 2015; Dekleva et al., 2018; Wise et al., 1997).The term premotor cortex was originally coined by Hines (Wise, 1985), owing to its adjacent position to and connectivity with motor cortex. This connectivity implies that the premotor cortex is hierarchically earlier than the motor cortex. Premotor cortex's major purpose is presumed to be to set the stage for motor cortex by potentiating certain motor plans and depotentiating other ones (Mirabella et al., 2011; Pastor-Bernier and Cisek, 2011).

The idea that premotor cortex regulates higher level motor plans rather than enacts them (e.g., by signaling spinal motor neurons) was first shown by Woolsey and colleagues (1952), who found stimulation of PMC did not produce movements (see also Travis, 1955). Later studies showed though that movements can be evoked by premotor stimulation, but are often more complex (e.g., whole hand grasping) those elicited by stimulation of primary motor cortex (e.g., specific muscle innervation; Graziano et al., 2002). The notion of premotor cortex as biasing downstream action execution is bolstered by its larger preoccupation during movement planning compared to online execution (Weinrich & Wise 1982). A high-level action regulatory role for premotor cortex is supported by recordings showing preparatory activity (Churchland et al., 2006) for reach direction specificity (Weinrich and Wise, 1982), encoding of multiple possible motor plans rather than a singular action (Cisek et al., 2006), switching between action plans, and during online control (Ames et al., 2019; Dekleva et al., 2018; Pastor-Bernier et al., 2012).

As a whole, these findings indicate that the premotor cortex has an ancillary motor function. That is, while it does not directly drive muscle specific responses, it plays an invaluable function: it sets the stage for action by making some actions more or less likely. In other words, it influences motor function by potentiating or depotentiating actions. In summary, then, the traditional view of the premotor cortex is that, anatomically and functionally, it resides at the first level of what we and others (reviewed next) argue is an action abstraction hierarchy. ***The essence of our argument is that other prefrontal regions extend this hierarchical control of action, and can also be described similarly to premotoric terms as (de)potentiating abstractions of action.*** In other words, other prefrontal regions do not differ from premotor cortex in kind, just in hierarchical level.

As we will argue below, the functions of these other regions, including their economic ones, can be explained, at least in part, by their premotor role. The major difference of PFC with premotor cortex is that they are anatomically and hierarchically earlier, and the actions they deal with are likely more abstract, more tangled, and more aligned with sensory input features than classically defined premotor cortex (Yoo and Hayden, 2018). At the extreme version of this perspective, one can see the entire brain as a kind of premotor structure. From this view, even retinas are premotor since high level form vision is there to drive choice and vision is just an untangled form of retinal processes. This view may seem to be true but trivially so; that is, it may cast a net so wide that it provides little in the way of novel guidance. Here we use the term premotor in a more restrictive sense. We believe that it is important to think of the prefrontal cortex as a premotor structure even if we aren't willing to think of the entire brain as a premotor structure. That is, within the broader action hierarchy, the more rostral regions that feed into the PFC can be given nameable functions, such as face detection, auditory localization, or gustatory identification. These structures may indeed be said to have a modular role.

We suspect the same is not likely to be true for the structures of the PFC. The prefrontal cortex has certain features, especially in its integration of information from multiple sources, that make it convenient to start there. We can simultaneously accept two points: (1) form vision is very useful for action selection and (2) evolution has apparently selected for a specialized dedicated visual system that serves the purpose of encoding visual form. This second point is critical. Apparently, it is a better design principle for visual inputs to converge and come to some consensus on form identity before integrating with other modalities, such as the visceral and olfactory systems. Regardless of the evolutionary reason, the visual system is conveniently thought of as a visual system – that is, it has a somewhat modular visual function. This is not to discount evidence for non-visual signals in the visual system, just to say that it has a strong bias towards visual function that other sensory cortical regions do not have. Other systems, for example, the brain's olfactory and auditory systems, may also have somewhat modular functions. The argument here is that PFC is not a modular system in the same way these ones are identifiable as modular. Instead, it reflects the convergence of multiple, more modular systems and serves as an important step in a hierarchy that produces action.

**Hierarchies of action: abstraction, control and goals**

In theorizing about the functions of the prefrontal cortex as premotoric, our view is anticipated by Joachin Fuster (Fuster, 2000, 2001, and 2015). Fuster viewed the prefrontal cortex as part of a hierarchy oriented towards the control of action. He said, for example, that “the entire cortex of the primate’s frontal lobe seems dedicated to organismic action. It can, thus, be considered, as a whole, ‘motor’ or ‘executive’ cortex in the broadest sense.” (Fuster, 2000). Note that Fuster here uses the word executive in the sense of executing action as distinguished from sensation, not in the sense of a discrete and separate executive or supervisory system.

Of course, Fuester does not mean that the entire prefrontal cortex is an undifferentiated mass of one extended premotor cortex. There are well described functional differences with the PFC, and there is a larger organization. To quote Fuster again, “much of the prevalent confusion in the PFC literature derives from two common errors. The first is to argue for one particular prefrontal function while opposing or neglecting others that complement it; the second is to localize any of them within a discrete portion of PFC.” (Fuster, 2001). In other words, Fuster proposes that the core function of the PFC is motor control, that its organization is hierarchical, and that its regions differ in their position, not in their nameable function.

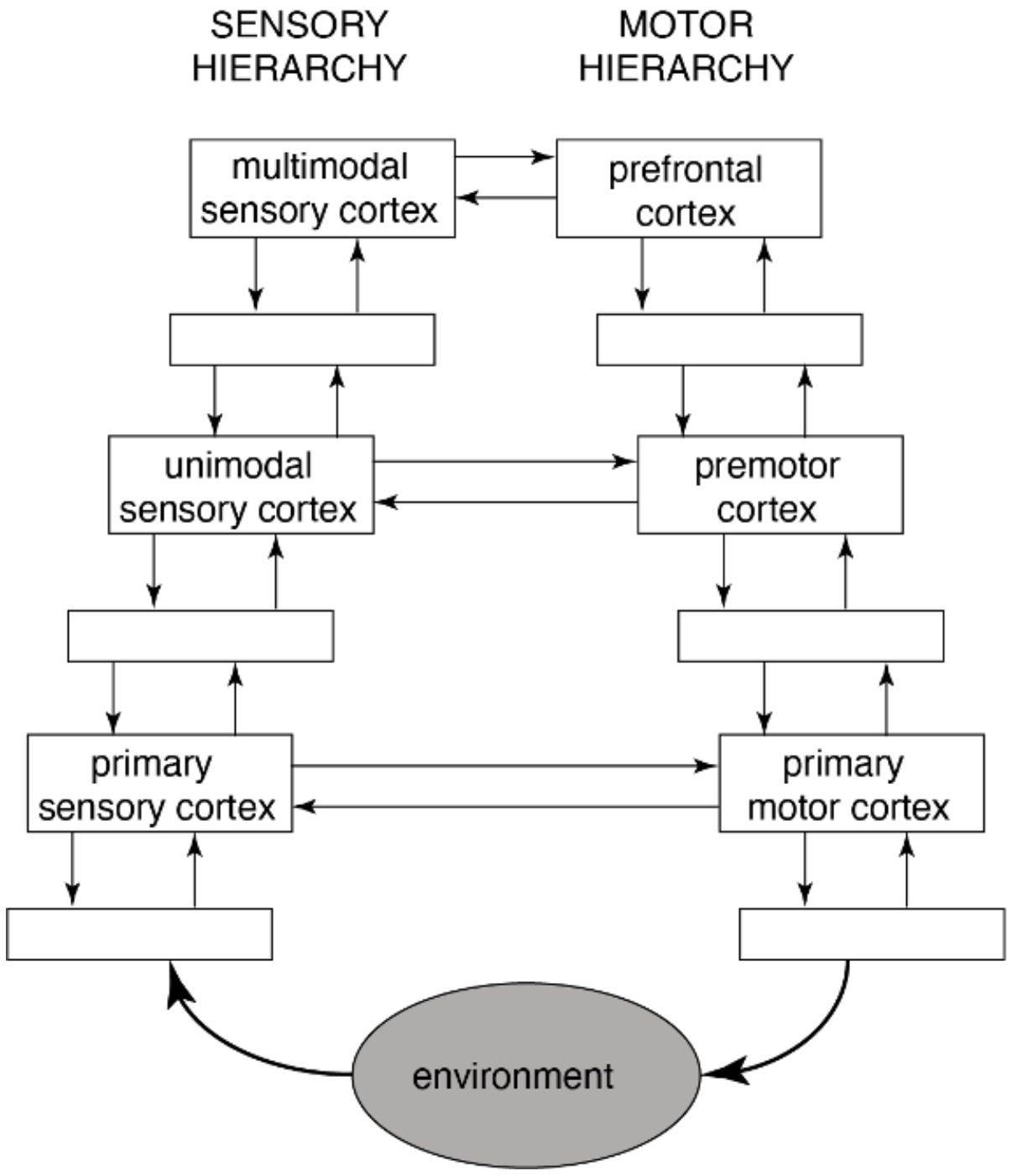

**Figure 1.** Functional organization of the prefrontal cortex, as proposed by Fuster (e.g. in 2000, 2001, and 2015). In this cartoon, the brain takes in information from the environment, processes it, and generates actions. The processing is hierarchical, and involves a gradual transformation from input to output. Critically, Fuster's proposal ignores a separate central executive.

Indeed, the notion of action and abstraction hierarchies already figures heavily into several theories of goal-directed cognitive control and decision-making. Generally, such hierarchical theories propose that areas within the prefrontal cortex can be described as some type of abstraction hierarchy over action control, culminating in the motor cortex and interacts with basal ganglia circuits (Badre, 2008; Badre and D'Esposito, 2009; Koechlin et al., 2003; Botvinick, 2008; Hunt et al., 2018; Hunt and Hayden, 2017). What differentiates these theories is how the brain deploys hierarchical abstraction to *control behavior* after a goal (e.g., feed your friends) is already specified.

This idea of hierarchical control can be illustrated by considering the case of a person who is interested in cooking dinner for a visitor. That goal could be satisfied by any number of possible actions, meaning that successful cooking of the food can be accomplished through any number of specific body movements. And indeed, the higher-level choice (what to cook) can be implemented in multiple steps (which things to cook in which order), and each of those can be executed in multiple specific actions (turn left to grab the saucepan, etc). So the decision about cooking is at a higher hierarchical level than the execution of the motor actions, although both are parts of control, broadly speaking. Indeed, planning and decision-making takes place at more levels than this - they involve a whole series of levels, including even more abstract ones, like whether cooking is best, or ordering takeout might be smarter.

Indeed, the ideas of abstraction and hierarchy for action already figure heavily in many theories of cognitive control and decision-making in PFC. Perhaps the most influential framework posits that abstraction hierarchies in PFC can be decomposed into different types of cognitive operations (Badre, 2007; Badre and Nee, 2018; Nee and D'esposito, 2016). These operations include abstractions over temporal information, schemas or states, and policy abstraction. Policy abstraction is most directly related to action control and aligns with the motor hierarchy in Futster's (2000) conception (Botvinick, 2007). Policies in this framework are rule-based mappings that are contextualized by more general rule-based policies. For example, having the goal state of entertaining a new person will map to the higher-order action of 'cooking for a date' is an example of policy or state-action mapping. A high-level goal like 'cooking' is an example of an abstract policy that generalizes over lower order policies constrained by contextual information regarding what to cook; context here could be a person's dietary preferences. In this example, policy abstraction links states and goals to potential action, with higher-order contexts like dietary preferences determining lower-order conditions for actions (e.g., make vegetarian not meat dish). Essentially, the least abstract policies during cooking a meal might directly map from states like 'in front of a heated stove' to an action of ‘grasp pan with right hand’. The states driving a that map policy actions are essentially a set of features containing information that affect the choice made, such as the person being cooked for combined with knowledge that they enjoy being cooked for is a state (Niv, 2019).

While these ideas of action hierarchy have enjoyed success in describing behavior, they have, however, primarily focused on how control is implemented through abstracted policies after goals are already specified. A formative question raised by these previous theories and others relates to how the goals that are directing behavior are themselves controlled or selected (Fine et al, 2020; O’Reilly et al., 2014). Building on this need to address the origin of goal selection, recent ideas on hierarchical action gradients have argued, for example, that either the OFC (including vmPFC; Holroyd and Verguts, 2021; O’Reilly, 2014) or frontal pole (Broadmann area 10) may be involved in either selecting, maintaining or distributing abstracted goal information (e.g., get food) to other cortical regions (Mansouri et al., 2017). However, these frameworks have largely considered how goals are maintained (e..g, working memory, Badre and Nee, 2018 ; O’Reilly, 2014), while the issue of the decision processes underlying goal-selection still awaits further elaboration.

In line with the above ideas on hierarchy and control, we propose that consideration of the homeostatic and motivational drives of behavior naturally leads to the idea that goal-selection is also an action decision-process that resides atop the PFC abstraction hierarchy. Notably, the motivations for our proposal closely align with Fuster’s conception of motor hierarchy and PFC (Fuster, 2000, 2001, and 2015). Key to our idea is treating goal-selection as a premotor action policy.

**Redefining goal selection as policy abstraction**

We now turn to delineating our theory in more concrete detail. We are proposing that PFC is, fundamentally, a *policy abstraction hierarchy*. That is, each area moving up the hierarchy has a map between states and progressively more abstract actions. Choice is not just “choosing left vs. right” right also choosing what goals to follow, *goal selection*. And, the difference between these is one of level, not of kind. The PFC as a whole serves the orchestration of goal-directed action (the actions one should take once the goal is specified) as well as the goal-selection (which goal is to be selected). While we take direct inspiration from the hierarchical theories reviewed in the previous section, the novel element of our proposal is that hierarchical abstraction of action policies be extended to include goal selection as the highest level of the process. We will argue, below, that this can be linked to the orbitofrontal cortex (OFC).

As an entry point to our idea, consider a descriptive example of a policy. For example, a person walking to work might come to a changing crosswalk light while simultaneously realizing they are already late to work. The individual could arbitrate between running across anyways or biding their time till the cross-light comes on again. All of these events and stimuli (*features*) constitute a *state*. Formally, states serve as a summary of all features that affect the choice of actions. Features could include the overall goal of going to work, relative time since the light changed and your distance, traffic conditions, or the emotional or physical cost of being late to work. The cost of being late could outweigh concerns for safety, for example, pushing an individual to cross. In this scenario, the policy is the mapping from how the states (the risk of getting hit, the cost of getting hit, the cost of being late) guide the person in deciding which action (cross or wait) to take.

Having an example of a policy at work, we can define the *policy* of our example as P(Action | State). The notation for the policy is read as the probability of an action conditioned on the state. The above example of street-crossing captures the idea that crossing or waiting is conditioned on the state and the goal of getting to work. This example accords well with what most theories mean by policy: action is almost always defined in terms of specific physical variables or action rules, e.g., reach or walk if the light is green but not red. Thus, while these policies operate in service of the goal of getting to work, the goal is already defined, and is merely an input that serves lower level actions.

In our view, though, our imagined person entertains policies that are richer and more abstract than rule-based action mappings that are subservient to a pre-specified goal of 'go to work'. For example, our person might have flexibility in their choice of daily schedules, such as the job allows them to decide whether to work or forgo it in favor of a pleasurable activity such as kayaking. In totality, we argue this higher-level choice exemplifies how the *action of selecting amongst (potentially competing) goals* is an abstract action policy, in line with the premotor notion of (de)potentiating different action.

A first step in defining what this goal selection policy would look like is defining the relevant state inputs (or, equivalently, motivators). (A very detailed overview of how motivators serve goal selection is found in O'Reilly, et al., 2014). Goal selection is molded by things a decision-maker wants or desires (e.g., money or pleasure), the external environment (e.g., opportunities during nice weather), pre-existing goals (e.g., work deadlines or dieting), or their homeostatic needs, such as hunger or thirst. These motivating states are often subject to depletion, meaning that a decision-maker must often prioritize goals that fulfill needs and balance resources based on their expected future depletion levels (Cannon, 1929; Keramati and Gutkin, 2014; Fine et al., 2020), and demarcate between wants and needs in goal-selection (O'Reilly, 2014; Juechems and Summerfield, 2019; Hull; 1943). Consequently, the evaluative processes underlying goal-selection dynamics are highly context-dependent, and driven by the needs for resource uptake (e.g., food or money), as well as desires (Fine et al., 2020; Keramati and Gutkin, 2014; Juechems et al., 2018; Juechems and Summerfield, 2019).

A major implication of the idea that motivation is often resource dependent is that optimizing goal selection policies requires an interaction with planning policies. In other words, goal selection must be future oriented. Anyone who has waited until they were hungry to go to a restaurant with a long-wait has experienced the consequences of failing to plan around these dynamics. One thing to keep in mind is that we likely do not pick a goal and then plan; instead, these things occurs simultaneously, and interactively (Fine et al., 2020; Pezzulo et al., 2019). The value of pursuing a particular goal (or multiple goals simultaneously) depends on the availability and feasibility of plans given different constraints. Interactions between planning processes and goal-selection dynamics are thus imperative for ensuring an agent ends up in future states where goals are met in a reasonable time, have trackable progress, and provide sufficient replenishment of resources (O'Reilly, 2014).

Planning, then, is another way in which goal selection is essentially premotor. To see this, we can compare the general algorithmic nature of planning for goal-selection and sensorimotor control (Todorov and Jordan, 2002). The two are identical algorithmically. Computationally, optimal planning of actions for achieving goals typically involves learning of a world model or connections between states and using knowledge of the states that satisfy goals to plan actions accordingly. For an agent to optimally plan while minimizing the distance and resources used, they must start planning from a goal satisfying state wherein they know needs or goals will be met, and move backwards to the person's current world state (Daw, 2012; Fine et al., 2020), e.g., imagining the path from work to home as taken by walking versus driving. This same algorithmic approach has been used to successfully describe sensorimotor control, such as planning motor dynamics to reach to a target (Shadmehr and Krakauer, 2008; Todorov and Jordan, 2002). These ideas indicate a conceptual and algorithmic overlap between planning for motor control policies and goal-selection policies, blurring the distinction between abstract goal-selection policies and sensorimotor control policies. As a proof of principle, these types of goal-selection dynamics have recently shown to be viable in a biologically realistic neural network (Fine et al., 2020) - fully local connections and hebbian learning.

These types of computations for policy optimization are often quite computationally complex - potentially beyond the limits of our brains to implement. Information-processing in the brain is inherently capacity-limited and noisy (Hick, 1952; Sims, 2016). Due to these informational constraints, goal selection cannot be optimized in an error-free manner (Bhui et al., 2021), due in part to a combinatorially high-dimensional state (and temporal) space that goals can be achieved in, uncertainty in whether pursuing a goal will render the desired outcomes, or the error-proneness of complex plans for goal achievement (Lai and Gershman, 2021). We suspect that understanding how individuals deal with these constraints will be necessary to elucidate the dynamics of goal-selection. This issue can be usefully reframed as asking how individuals' trade-offs between the complexity of a goal-selection policy and plans with the potential benefits or needs of satisfying certain goals. Normative solutions to these problems include bounded rationality or rate-distortion frameworks, both of which have already been applied to understand action policy optimization and decision-making (Lai and Gershman, 2021;

Genewein et al., 2015). In short, these frameworks suggest policies are encoded to minimize redundancy and be efficient by compressing the policy encoding in the brain.

Several testable predictions emerge when applying these frameworks to goal-selection policies, two of which we consider here. A notable prediction is that goal-selection should exhibit a trade-off between (1) the urgency, amount, and quality of resources gained to fulfill a need and (2) the temporal (or distance) and the state-space complexity of a plan for achieving it. This distinction has been experienced by anyone who knows cooking something would be healthier than ordering delivery, but cooking is more state-complex than picking up the phone to order and wait for delivery. Another prediction is that the imperative for compressing the policy is that individuals can learn to abstract over goal-fulfilling states. For example, our person walking to work may represent the world at different level of abstraction depending on need and desire - if thirsty, they may classify shops into ones that can provide a drink or not; if thirsty and hungry, they may instead classify shops into ones offering drinks and food or not. The extent of goal-fulfilling state abstraction should play a direct role in whether goals are separated or merged. The point we want to convey in discussing this framework and examples is that goal-selection policies are not merely an abstract thought exercise. They represent a plausible component of action abstraction hierarchies, fit the notion of premotor, and have empirically testable predictions that are grounded in extant theories of optimizing decision-making (Lai and Gershman, 2021; Genewein et al., 2015; Todorov and Jordan, 2002).

**Reconsidering the role of OFC as residing atop the premotor PFC hierarchy**

To examine how these ideas play out in neurophysiology, we now turn to considering one brain area, the orbitofrontal cortex (OFC), the presumed start of the premotor hierarchy and core component of goal-selection. The OFC sits immediately superior to the orbits of the eyes (Wallis, 2007; Schoenbaum et al., 2009; Rudebeck and Murray, 2014; Rudebeck and Rich, 2018). In the primate (including humans), it consists of areas 11 and 13; sometimes other adjacent areas are included (Carmichael and Price, 1994). Understanding OFC function is critical because of, among other reasons, its central role in important psychiatric diseases, including addiction and depression (e.g. Drevets, 2007; Volkow and Fowler, 2000). More broadly, OFC plays a role in basic decision-making and is a wellspring for motivation, volition, and, ultimately, free choice (Wallis, 2007; Padoa-Schioppa, 2011; Murray and Rudebeck, 2018; Plassmann et al., 2007; Price, 2005; Soon et al., 2013).

We are proposing that the canonical function of OFC is optimization of goal selection policies. This proposal is consistent with several recent models and neurophysiological findings that portray OFC as performing goal-selection and representing the distance to achieving a goal-fulfilling state (Castegnetti et al., 2021; Fine et al., 2020; Juechems et al., 2018; Zarr and Brown, 2021). OFC is well positioned to be at the top and most abstract portion of the prefrontal hierarchy. One factor in favor of this idea is that OFC has a somewhat unique anatomy (Rudebeck and Murray, 2014; Rushworth et al., 2011). It receives inputs from a diverse array of regions with heavily specialized functions, positioning OFC as a hub for integrating disparate information sources and forming inferences. These connections to OFC include four of the five major senses (all except the auditory system), from visceral areas, from hippocampus and amygdala, and from the ventral striatum (Ongur and Price, 2000; Haber and Knutson, 2010; Price, 2005; Carmichael and Price, 1995a, Carmichael and Price, 1995b, Ghashghaei et al., 2007). It does not have direct access to motor or premotor regions. However, its ability to influence them indirectly is clear. For example, it has direct projections to the ventral, medial, and dorsolateral PFC (Barbas and Pandya, 1989, Carmichael and Price, 1996, Saleem et al., 2014), which allow it indirect descending control over dorsal premotor areas. That is, it is possible to place it at the apex of a series of regions that, in a chain, influence the next in the series, to ultimately drive the motor cortex and other regions with direct spinal motorneuron access. What distinguishes OFC from other PFC areas is that it's the first gathering point for distinct and relatively discrete sensory and association streams.

Our theory, which portrays OFC as a premotor structure that optimizes goal-selection, contrasts with the well-known theory that OFC is predominantly an economic structure (Padoa-Schioppa, 2011; Wallis, 2007; Schoenbaum et al., 2011; Wallis, 2012; O'Doherty, 2004; Rushworth, 2011). The economics view emphasizes the contributions of OFC to evaluating options and for comparing values to select a preferred one. While this view has undoubted validity, it has three limitations. *First*, it is not clear to what extent the OFC is more economic than other brain regions. Indeed, a good deal of evidence supports the idea that economic representations are highly distributed, and that comparisons reflect the outcome of processes occurring in multiple brain regions (Cisek, 2012; Hunt and Hayden, 2017; Vickery et al., 2011; Hunt et al., 2014). *Second*, it is not clear the extent to which OFC shows specialization for economic functions. That is, OFC appears to participate in many cognitive processes, including those that are only indirectly related to economic decision-making. For example, research implicates OFC in representation of sensory details of predictions (e.g. Burke et al., 2008, Tsujimoto et al., 2011 and 2012), of abstract rules (Sleezer et al., 2016; Wallis et al., 2001; Buckley et al., 2009), and task- and state-switching (Sleezer et al., 2017; Tsujimoto et al., 2011; Young and Shapiro, 2009; Gremel and Costa, 2013). *Third*, and most important, OFC's apparent value coding appears, on closer inspection, to reflect expectancy signalling rather than value coding (Jones et al., 2012; Zhou et al., 2021a and b; Gardner and Schoenbaum, 2021; Wang and Hayden, 2017; Farovik et al., 2015).

The non-economic view has reached its greatest level of sophistication in the cognitive map of task space theory (Niv 2019; Schuck et al., 2016, Wilson et al., 2014; Wikenheiser and Schoenbaum, 2016). That is, its responses serve to encode the set of relevant mappings associated with potential actions and options in the current environment. As such it serves as a potential source of information that can guide decision-making and action selection. This set of mappings may serve as a superset of encodings that also includes reward information, meaning OFC may be more than just an economic predictor (Niv, 2019; Wilson et al., 2014; Wikenheiser and Schoenbaum, 2016; Rudebeck and Murray, 2018). It may also explain, for example, rule encoding in OFC (Sleezer et al., 2016; Wallis et al., 2001). That should also include information about space. Indeed, the encoding of spatial information has taken on an important position in debates about the mechanisms of choice and valuation in the orbitofrontal cortex (Padoa-Schioppa, 2011; Grattan and Glimcher, 2014; Strait et al., 2016; Yoo et al., 2018). Put differently, evidence for a lack of spatial selectivity would support the notion of a modular, non-premotor, purely economic OFC. Despite the debate, a large set of evidence demonstrates spatial selectivity within OFC. Spatial selectivity is observed in neurons in primates (Tsujimoto et al., 2009; Abe and Lee, 2011; Luk and Wallis, 2013; Strait et al., 2016; McGinty et al., 2016; Yoo et al., 2018; Roesch and Olson, 2004; Baeg et al., 2020; Costa and Averbeck, 2020) and in rodents (Roesch et al., 2006; Feierstein et al., 2006; Furuyashiki et al., 2008; Sul et al., 2010; Van Wingerden et al., 2010; Bryden Roesch 2015; Young and Shapiro, 2011). Considered together, this work clearly indicates that single neurons in OFC respond with information about the spatial details of the task at hand. We suspect that OFC's capacity for encoding of spatial information is particularly important, as this type of information is necessary for any goal-selecting organism to decide upon the usefulness of a future goal-fulfilling state. Linking these views, these findings suggest that OFC's value representations are one of a much broader set of functions that involve integrating information useful for guiding decision-making over goals and behavior. In that view, then, the apparent economic functions of OFC are a consequence of the fact that value is one of many factors that influence abstract action or goal selection.

**If the prefrontal cortex is premotor, we need to use dynamical systems approaches to analyze it**

We have proposed that the entire prefrontal cortex can be thought of as a premotor structure. If that is true, then how best to analyze its activity to understand its computations? We propose that the best analytical tools for studying prefrontal cortex are those applied to studying motor and premotor cortex (as it is normally defined). Specifically, important theoretical and empirical scholarship has challenged the long-dominant representational theory of motor activity, in which firing rates of single neurons reflect potentiation of specific action related quantities, such as velocity or force (Churchland, 2010; Kaufman et al., 2014; Shenoy et al., 2013; Vyas et al., 2020). In the dynamical systems framework, responses of single neurons are a projection of a computation occurring in a higher dimensional space implemented by populations of neurons.

Indeed, we believe that the dynamical systems revolution has not gone far enough (Ebitz and Hayden, 2021; see also Urai et al., 2021). We believe the focus of how populations and their ensuing dynamics implement behaviorally relevant computations should be extended beyond the motor cortex, to the entirety of the prefrontal cortex. Importantly, while the traditional idea of isomorphism between neural and psychological variables is generally rejected by adherents of the dynamical systems perspective, it does actually do a decent job accounting for firing rates of neurons - but that is not evidence that those neurons serve to represent those variables (Fetz, 1992; Churchland and Shenoy, 2007). Likewise, years of results showing correlations between firing rates of neurons in prefrontal areas and task variables does not constitute evidence for the representationalist hypothesis in PFC.

The dynamical systems hypothesis reinforces the importance of studying population, and how single units contribute to the ensemble-wide computation. The firing rate of any given neuron is the projection of the ensemble activity (Saxena and Cunningham, 2019). Insights into neural activity require inferring the properties of the system as a whole, which typically requires examining the properties of populations of neurons and theories linking how dynamics properties such as line attractors, for example, can implement computations unders study (e.g. information integration or gating, Mante et al., 2013). The dynamical systems approach offers potential ready solutions to problems that are difficult to solve using the representational paradigm. A good example of this is the need to functionally partition motor planning from execution (Elsayed et al., 2016;  Kaufman et al., 2014). We do not simply act - we must prepare our actions, and that preparation involves the same circuits, indeed, the same neurons, as are used during execution. This functional overlap between planning and execution raises an engineering problem - how to prevent precocious movement. Proposed solutions, such as subthreshold firing, are not empirically supported. The dynamical systems perspective offers a simple solution - the computation can be made in an orthogonal task-null subspace, and then rotated at the time of execution. This solution makes use of the full dynamic range of the neurons and is supported by empirical findings. Notably, the need to prevent precocious responding is also a potential problem if we see the PFC as a premotor, and data supports the idea that the same orthogonalization process is used in earlier areas as well (Yoo and Hayden, 2020).

**Conclusion**

We propose that the prefrontal cortex is, in essence, a hierarchically organized premotor structure. Casting its organization this way can help to organize our understanding of its activities. This view, then, sees the prefrontal cortex as the mirror reversed complement of the sensory systems, especially the ventral visual system, which is organized along the axis of ever more complex form representation (Yoo and Hayden, 2018). We propose that viewing the prefrontal cortex in this manner will help resolve important debates, and will push researchers away from the quest of identifying the "essential function" of each region within it, and instead to understanding how it coordinates its computations to produce action.